\documentclass[%
 reprint,
superscriptaddress,
preprintnumbers,
nofootinbib,
 amsmath,amssymb,
 aps,
]{revtex4-2}

\usepackage{graphicx}
\usepackage{dcolumn}
\usepackage{bm}
\usepackage{hyperref}

\hypersetup{
    unicode=false,          
    pdftoolbar=true,        
    pdfmenubar=true,        
    pdffitwindow=false,     
    pdfstartview={FitH},    
    pdftitle={},    
    pdfauthor={},     
    pdfsubject={},   
    pdfcreator={},   
    pdfproducer={}, 
    pdfkeywords={} {} {}, 
    pdfnewwindow=true,      
    colorlinks=true,       
    linkcolor=blue, 
    citecolor=blue,        
    filecolor=magenta,      
    urlcolor=blue,           
	breaklinks=true
} 

\usepackage{amssymb}
\usepackage{amsmath}
\usepackage{amsfonts}
\usepackage{xspace}
\usepackage{bm,bbm}
\usepackage{relsize}
\usepackage{siunitx}
\usepackage{xcolor}
\usepackage{upgreek}

\makeatletter
\DeclareFontFamily{OMX}{MnSymbolE}{}
\DeclareSymbolFont{MnLargeSymbols}{OMX}{MnSymbolE}{m}{n}
\SetSymbolFont{MnLargeSymbols}{bold}{OMX}{MnSymbolE}{b}{n}
\DeclareFontShape{OMX}{MnSymbolE}{m}{n}{
    <-6>  MnSymbolE5
   <6-7>  MnSymbolE6
   <7-8>  MnSymbolE7
   <8-9>  MnSymbolE8
   <9-10> MnSymbolE9
  <10-12> MnSymbolE10
  <12->   MnSymbolE12
}{}
\DeclareFontShape{OMX}{MnSymbolE}{b}{n}{
    <-6>  MnSymbolE-Bold5
   <6-7>  MnSymbolE-Bold6
   <7-8>  MnSymbolE-Bold7
   <8-9>  MnSymbolE-Bold8
   <9-10> MnSymbolE-Bold9
  <10-12> MnSymbolE-Bold10
  <12->   MnSymbolE-Bold12
}{}

\let\llangle\@undefined
\let\rrangle\@undefined
\DeclareMathDelimiter{\llangle}{\mathopen}%
                     {MnLargeSymbols}{'164}{MnLargeSymbols}{'164}
\DeclareMathDelimiter{\rrangle}{\mathclose}%
                     {MnLargeSymbols}{'171}{MnLargeSymbols}{'171}
\makeatother

\newcommand{\norm}[1]{\lVert#1\rVert}

\begin{document}

\title{Probing topology in nonlinear topological materials using numerical $K$-theory}

\author{Stephan Wong}
\email[Email: ]{stewong@sandia.gov}
\affiliation{Center for Integrated Nanotechnologies, Sandia National Laboratories, Albuquerque, New Mexico 87185, USA}

\author{Terry A. Loring}
\affiliation{Department of Mathematics and Statistics, University of New Mexico, Albuquerque, New Mexico 87131, USA}

\author{Alexander Cerjan}
\affiliation{Center for Integrated Nanotechnologies, Sandia National Laboratories, Albuquerque, New Mexico 87185, USA}

\date{\today}

\begin{abstract}
Nonlinear topological insulators have garnered substantial recent attention as they have both enabled the discovery of new physics due to interparticle interactions, and may have applications in photonic devices such as topological lasers and frequency combs.
However, due to the local nature of nonlinearities, previous attempts to classify the topology of nonlinear systems have required significant approximations that must be tailored to individual systems.
Here, we develop a general framework for classifying the topology of nonlinear materials in any discrete symmetry class and any physical dimension. Our approach is rooted in a numerical $K$-theoretic method called the spectral localizer, which leverages a real-space perspective of a system to define local topological markers and a local measure of topological protection. Our nonlinear spectral localizer framework yields a quantitative definition of topologically non-trivial nonlinear modes that are distinguished by the appearance of a topological interface surrounding the mode. Moreover, we show how the nonlinear spectral localizer can be used to understand a system's topological dynamics, i.e., the time-evolution of nonlinearly induced topological domains within a system. We anticipate that this framework will enable the discovery and development of novel topological systems across a broad range of nonlinear materials.
\end{abstract}

\maketitle


\section{Introduction}

Over the past two decades, the discovery of topological materials has revolutionized a broad range of materials research through the prediction and observation of fundamentally new classes of states that are robust against defects and imperfections~\cite{Chiu2016, Ozawa2019, Ma2019, Cooper2019}.
In non-interacting systems, the possible types of topology that a material can exhibit are determined by its discrete symmetries (or lack-thereof), yielding a periodic table of material topology rooted in the ten Altland-Zirnbauer classes~\cite{Altland1997, Schnyder2008, Kitaev2009, Ryu2010}.
This topological classification framework has subsequently been expanded to include crystalline symmetries~\cite{Ando2015} and semimetals~\cite{Yan2017, Burkov2011}.
For all of these cases, a non-interacting material's topology is traditionally determined through the calculation of invariants built from the system's band structure and Bloch eigenstates, and thus the topological invariants are global properties of the bulk material.

However, in many systems, interactions are both unavoidable and potentially highly desirable, as they can result in emergent phenomena~\cite{Chiu2016, Rachel2018, Ozawa2019, Cooper2019}. For example, topologically ordered phases of matter can support quasiparticles with fractional charge and anyonic statistics~\cite{Nayak2008, Stern2008}.
More recently, studies of interacting bosons in the mean-field limit have led to the discovery of new physics~\cite{Smirnova2020}, 
such as the appearance of topological bulk~\cite{Lumer2013, Marzuola2019, Mukherjee2020, Jurgensen2021, Li2022, Li2022a, Jurgensen2022, Ren2023, Jurgensen2023}
and edge solitons~\cite{Leykam2016, Smirnova2019, Mukherjee2021}, the observation of nonlinearly induced topological phase transitions~\cite{Hadad2016, Zhou2017, Chaunsali2019}, and the concept of multi-wave mixing with topological states~\cite{Pilozzi2017, Zhang2019, Mittal2021, Jia2023}. Such bosonic systems are usually described using the nonlinear Schr{\" o}dinger equation, in which the nonlinear response is parameterized by the strength of the inter-particle interactions. Unfortunately, nonlinearities present a substantial challenge for our understanding of material topology as they demand a shift in perspective. Whereas non-interacting (linear) systems are topologically classified using only a system's single-particle Hamiltonian, classifying the topology of a nonlinear system requires knowing its occupation. Moreover, as nonlinear effects are intrinsically local, they can break the fundamental assumption of band theory (which demands spatial periodicity). Altogether, these challenges have so far prohibited the development of a broadly applicable theory for classifying the topology of nonlinear systems and their possible occupations.



%

In this paper, we present a general framework for probing the topology of nonlinear materials based on numerical $K$-theory. 
Our framework classifies a system's topology in real-space using local markers, and as such directly accommodates spatial inhomogeneities in the system's occupation and the material's response. 
To do so, we first determine the system's Hamiltonian accounting for its occupation, and then combine it with the system's position operators using a non-trivial Clifford representation. 
This yields the system's spectral localizer, which defines local markers for all ten Altland-Zirnbauer classes in any physical dimension, as well as a local measure of topological protection~\cite{Loring2015}.
Using this classification framework, we develop a rigorous definition of topological nonlinear modes distinguished by the appearance of nonlinearity-induced topological interfaces surrounding them; the topological robustness of these nonlinear states guarantees their existence over some range of finite perturbations (i.e., a solution to the nonlinear Hamiltonian is guaranteed).
Moreover, our framework allows for probing topological dynamics, and one can numerically observe the transport of a nonlinearity-induced topological phase with the propagation of the nonlinear state that creates it.
%
%
%
%
%
Overall, our classification framework provides a general approach for the characterization of stationary and dynamical topological phenomena in nonlinear materials in any discrete symmetry class and in any physical dimension. 
We anticipate that our nonlinear spectral localizer framework will prove useful for the design and development of a broad range of novel nonlinear topological phenomena such as in pump-probe-like systems and topological frequency combs, which require simultaneous local information at different positions and energies.


\section{Results}


\subsection{Spectral localizer for nonlinear materials}
\label{sect:loclaizer}

In the past few years, a local, real-space approach for classifying non-interacting topological materials has been developed based on recent discoveries from the study of (possibly real, possibly graded) $C^*$-algebras~\cite{Loring2015, Loring2017, Loring2020}. 
The overarching idea of this approach is to combine a system's single-particle Hamiltonian with information about its real-space structure to form a single Hermitian composite operator called the \textit{spectral localizer}. 
The system's topology at a specified location and energy can then be determined using the original invariants proposed by Kitaev for 0D and 1D systems (i.e., the partitioning a system's spectrum about some gap for $\mathbb{Z}$ invariants, or signs of determinants or signs of Pfaffians of some portion of a system's Hamiltonian for $\mathbb{Z}_2$ invariants)~\cite{Kitaev2006}, but applied to the spectral localizer instead of the system's Hamiltonian. 
In other words, the spectral localizer approach is performing dimensional reduction consistent with Bott periodicity, such that the invariants of the fictitious dimensionally reduced system determine the local topology of the original system.

Our general framework for classifying the topology of nonlinear systems is built on the spectral localizer. 
The key advantage of this approach is that as the spectral localizer incorporates information about a system's spatial configuration to produce a real-space theory of material topology, it can be augmented to include the local nature of nonlinear effects.
Consider a $d$-dimensional nonlinear Hermitian system characterized by the nonlinear eigenvalue equation 
\begin{equation}
\label{eq:nl_eq}
H_\text{NL}(\boldsymbol{\uppsi}_\text{NL}) \boldsymbol{\uppsi}_\text{NL} = E_\text{NL} \boldsymbol{\uppsi}_\text{NL},
\end{equation}
with $\boldsymbol{\uppsi}_\text{NL}$ and $E_\text{NL}$ being the nonlinear eigenmode and its associated nonlinear eigenenergy, respectively. 
For such a system, the nonlinear spectral localizer $L_{\boldsymbol{\uplambda}}$ is a Hermitian matrix that combines the system's nonlinear Hamiltonian accounting for its current occupation $\boldsymbol{\uppsi}$ with its position operators $X_1, \dots, X_d$ using a non-trivial Clifford representation,
\begin{multline}
\label{eq:dLoc}
L_{\boldsymbol{\uplambda} = (x_1,\dots,x_d,E)}(X_1, \dots, X_d, H_\text{NL}(\boldsymbol{\uppsi})) = \\
\sum_{j=1}^d \kappa (X_j - x_j I) \otimes \Gamma_j + (H_\text{NL}(\boldsymbol{\uppsi}) - EI) \otimes \Gamma_{d+1}
. 
\end{multline}
Here, $\Gamma_1,\dots,\Gamma_{d+1}$ form a $(d+1)$-dimensional Clifford representation and satisfy $\Gamma_j^\dagger = \Gamma_j$, $\Gamma_j^2 = I$, and $\Gamma_j \Gamma_l = -\Gamma_l \Gamma_j$ for $j \ne l$, while $I$ is the identity matrix.
In a typical tight-binding basis, the position matrices $X_j$ are diagonal matrices where the $n$-th entry corresponds to the $j$-th real-space coordinate $(x_1^{(n)}, \dots, x_j^{(n)}, \dots, x_d^{(n)})$ of the $n$-th lattice site, namely 
\begin{equation}
X_j = 
\left(
\begin{array}{ccc}
\ddots &  &  \\ 
 & x_j^{(n)} &  \\ 
 &  & \ddots
\end{array} 
\right)
.
\end{equation}
Finally, in Eq.~\eqref{eq:dLoc}, $\boldsymbol{\uplambda} = (x_1, \dots, x_d,E)$ is to be seen as an input for locally probing the topology in real-space at the spatial coordinate $(x_1, \dots, x_d)$ and energy $E$, and $\kappa$ is a hyperparameter chosen to make the units consistent between the position and Hamiltonian matrices. 
The hyperparameter $\kappa$ is also set to balance the spectral emphasis on the system's position information relative to its Hamiltonian, and has been proven in non-interacting systems to have a broad range of applicability in topological insulators~\cite{Loring2017}. 
Moreover, $\kappa$ has been numerically observed to have utility beyond this limit~\cite{Loring2015, Cerjan2022, Cerjan2022a}.

Unlike other approaches to material topology that rely upon knowing a system's exact spectrum and associated single-particle energy eigenstates, the nonlinear spectral localizer is a multi-operator pseudospectral method, and returns information about a system's approximate joint spectrum across the non-commuting operators $X_1, \dots, X_d$, and $H_\text{NL}(\boldsymbol{\uppsi})$. 
For a given choice of $\boldsymbol{\uplambda} = (x_1, \dots, x_d, E)$, the spectrum of the nonlinear spectral localizer $\sigma(L_{\boldsymbol{\uplambda}})$ not only returns a measure of whether the system linearized about its current occupation supports a state $\boldsymbol{\upphi}$ approximately localized across all of $X_1, \dots, X_d$, and $H_\text{NL}(\boldsymbol{\uppsi})$ (i.e., such that $X_j \boldsymbol{\upphi} \approx x_j \boldsymbol{\upphi}$ and $H_\text{NL}(\boldsymbol{\uppsi}) \boldsymbol{\upphi} \approx E \boldsymbol{\upphi}$), but also information about how large of a perturbation is necessary to relocate one of the linearized system's states to be approximately localized at $(x_1, \dots, x_d, E)$.
The measure of this required perturbation is given by the smallest singular value of $L_{\boldsymbol{\uplambda}}$, i.e.,
\begin{multline}
\label{eq:localizer_gap}
\mu^\text{C}_{\boldsymbol{\uplambda}}(X_1, \dots, X_d,H_\text{NL}(\boldsymbol{\uppsi})) = \\ 
\min \big[ | \sigma(L_{\boldsymbol{\uplambda}}(X_1, \dots, X_d,H_\text{NL}(\boldsymbol{\uppsi}))) | \big]
.
\end{multline}
Small values of $\mu^\text{C}_{\boldsymbol{\uplambda}}$ indicate the existence of such a joint approximate eigenvector for $X_1, \dots, X_d$, and $H_\text{NL}(\boldsymbol{\uppsi})$ localized near $\boldsymbol{\uplambda}$, while large values of $\mu^\text{C}_{\boldsymbol{\uplambda}}$ indicate that the system does not exhibit such a state.
As such, $\mu^\text{C}_{\boldsymbol{\uplambda}}$ can be thought of as a local band gap. 
In Eq.~\eqref{eq:localizer_gap}, the superscript $\textrm{C}$ denotes Clifford, as the system's Clifford $\epsilon$-pseudospectrum is defined by those $\boldsymbol{\uplambda}$ where $\mu^\text{C}_{\boldsymbol{\uplambda}} \le \epsilon$~\cite{Loring2015,Cerjan2023a}.

By directly incorporating the nonlinear system's current occupation, the nonlinear spectral localizer can be used to classify the topology of systems in any discrete symmetry class (i.e.\ the ten Altland-Zirnbauer classes \cite{Chiu2016}) and any physical dimension by leveraging the corresponding local markers known for non-interacting systems~\cite{Loring2015}. 
Here, we focus on systems with the possibility of being Chern insulators (i.e., 2D class A systems), as the preponderance of studies of nonlinear topological materials have considered such systems~\cite{Lumer2013, Leykam2016, Mukherjee2021, Jurgensen2021, Jurgensen2022, Marzuola2019, Zhang2019, Mittal2021, Li2022a, Li2022}.
To assess whether a nonlinear 2D system possesses spatial regions and energy gaps where it is a Chern insulator using the spectral localizer, the Pauli spin matrices can be used as the non-trivial Clifford representation in Eq.~\eqref{eq:dLoc}, allowing it to be rewritten as
\begin{align}
\label{eq:localizer}
\begin{split}
& L_{\boldsymbol{\uplambda}=(x,y,E)}(X, Y, H_\text{NL}(\boldsymbol{\uppsi})) = \\
& 
\left(
\begin{array}{cc}
H_\text{NL}(\boldsymbol{\uppsi})-EI & \kappa (X-xI) - i\kappa(Y-yI) \\ 
\kappa (X-xI) + i\kappa(Y-yI) & -(H_\text{NL}(\boldsymbol{\uppsi})-EI)
\end{array}
\right)
.
\end{split}
\end{align}
Then, the system's local topology can be classified using the index
\begin{multline}
\label{eq:local_chern_nb}
C_{(x,y,E)}^{\textrm{L}}(X,Y,H_\text{NL}(\boldsymbol{\uppsi})) = \\
\frac{1}{2} \textrm{sig} \big[ L_{(x,y,E)}(X,Y,H_\text{NL}(\boldsymbol{\uppsi})) \big]
,
\end{multline}
where $\textrm{sig}(M)$ is the signature of the matrix $M$, i.e., its number of positive eigenvalues minus its number of negative ones. 
Intuitively, the spectral localizer is projecting the 2D system into 0D via the choice of $\boldsymbol{\uplambda} = (x,y,E)$, with the ``Hamiltonian'' of this dimensionally reduced system being $L_{\boldsymbol{\uplambda}}$. 
Then, the topology of the effective 0D system is determined through the partitioning of $L_{\boldsymbol{\lambda}}$'s spectrum about zero. 
This definition corresponds to a local Chern marker because it does not depend on the system possessing any discrete symmetries, and it is provably equivalent to the global Chern number for infinite, linear, crystalline systems~\cite{Loring2020}.

Together, the spectral localizer's local gap, Eq.~\eqref{eq:localizer_gap}, and local topological markers, such as Eq.~\eqref{eq:local_chern_nb}, form a consistent and complete picture of material topology. 
In particular, the measure of a system's topological robustness at a given location and energy is $\mu^\text{C}_{\boldsymbol{\uplambda}}$, because the spectral localizer's local markers are all well-behaved functions of its spectrum, and cannot change without $\mu^\text{C}_{\boldsymbol{\uplambda}} \rightarrow 0$. 
For example, for $L_{\boldsymbol{\uplambda}}$'s signature to change, one of its eigenvalues must cross $0$, at which point $\mu^\text{C}_{\boldsymbol{\uplambda}} = 0$. 
Likewise, locations where $\mu^\text{C}_{\boldsymbol{\uplambda}} = 0$ indicate interfaces where a system's local topology can change, and also require the system to exhibit a state approximately localized at $\boldsymbol{\uplambda}$, thus realizing bulk-boundary correspondence.

As topological protection in the nonlinear spectral localizer is determined by its smallest singular value, and as a Hermitian operator it has Lipschitz continuous singular values, the spectral localizer approach can be used to make rigorous predictions about the effects of a perturbation or defect. 
If $\delta H$ is a Hermitian perturbation distribution to the underlying system and $w$ is the perturbation strength, then the perturbed nonlinear Hamiltonian is
\begin{equation}
\label{eq:H_pert}
H_\text{pert}(w) = H_\text{NL}(\boldsymbol{\uppsi}_w) + w \delta H
,
\end{equation}
where $\boldsymbol{\uppsi}_w$ is the nonlinear eigenvector of $H_\text{pert}(w)$, and the change in the full Hamiltonian accounting for the perturbation's effects on the system's nonlinear eigenmode is 
\begin{equation}
\Delta H(w) = H_\text{pert}(w) - H_\text{pert}(0)
.
\end{equation}
Weyl's Perturbation Theorem~\cite{Bhatia1997} guarantees that the change of the eigenvalues of a Hermitian matrix $A$ relative to the eigenvalues of another Hermitian matrix $B$ is bounded by $\norm{A-B}$, i.e., the largest singular value of $A-B$. 
Therefore, when a perturbation is added to a system, the resulting change to its local gap is bounded (see also Ref.~\cite{Loring2015}, Lemma 7.2, for the linear case),
\begin{multline}
| \mu^\text{C}_{\boldsymbol{\lambda}}(\mathbf{X},H_\text{pert}(w)) - \mu^\text{C}_{\boldsymbol{\lambda}}(\mathbf{X},H_\text{pert}(0)) | \le \\
\norm{L_{\boldsymbol{\uplambda}}(\mathbf{X}, H_\text{pert}(w)) - L_{\boldsymbol{\uplambda}}(\mathbf{X}, H_\text{pert}(0))}
,
\end{multline}
where $\mathbf{X} = (X_1, \dots, X_d)$. Moreover, as the perturbation is only changing the system's Hamiltonian, and not its position operators,
\begin{multline}
\norm{L_{\boldsymbol{\uplambda}}(\mathbf{X}, H_\text{pert}(w)) - L_{\boldsymbol{\uplambda}}(\mathbf{X}, H_\text{pert}(0))} = \\
\norm{H_\text{pert}(w) - H_\text{pert}(0)} 
.
\end{multline}
Thus, the change in the local gap is bounded by the change of the nonlinear Hamiltonian,
\begin{multline}
| \mu^\text{C}_{\boldsymbol{\lambda}}(\mathbf{X},H_\text{pert}(w)) - \mu^\text{C}_{\boldsymbol{\lambda}}(\mathbf{X},H_\text{pert}(0)) | \le \\
\norm{\Delta H(w)}
.
\end{multline}
Altogether, this argument proves that a perturbation cannot change the system's topology at a given location and energy if the change in the nonlinear Hamiltonian $\norm{\Delta H(w)}$ is less than the local gap of the unperturbed nonlinear system $\mu^\text{C}_{\boldsymbol{\lambda}}(\mathbf{X},H_\text{pert}(0))$, as the minimum perturbation necessary to force the system to a topological phase transition has $\mu^\text{C}_{\boldsymbol{\lambda}}(\mathbf{X},H_\text{pert}(w)) = 0$. 
In other words, for a system that may be a Chern insulator,
\begin{multline}
\norm{\Delta H(w)} \le \mu^\text{C}_{\boldsymbol{\lambda}}(X,Y,H_\text{pert}(0)) \implies \\ C_{\boldsymbol{\uplambda}}^{\textrm{L}}(X,Y,H_\text{pert}(w)) = C_{\boldsymbol{\uplambda}}^{\textrm{L}}(X,Y,H_\text{pert}(0))
.
\end{multline}
As such, given knowledge of an ordered system's spectral localizer, one can immediately determine whether a given perturbation can cause a topological transition through a numerically efficient calculation.

This argument of topological protection also guarantees that the topology of a nonlinear system can be used in a pump-probe configuration; any sufficiently weak probe can be considered as a perturbation to a strong pump occupation present in the system, and this will not immediately change the system's local topology. In particular, this allows for any nonlinearly induced topological interfaces to support weak boundary-localized states that must exist due to bulk-boundary correspondence. However, the introduction of a weak probe signal can yield a decay time for the system's topology, as the interaction between the two excitations can slowly cause the pump to delocalize, eventually dissipating the system's nonlinearly induced topology.


\subsection{Topological nonlinear states}

Using the nonlinear spectral localizer, we can construct a rigorous definition for topological nonlinear states: 
A nonlinear eigenstate $\boldsymbol{\uppsi}_\text{NL}$ is topological if it creates a change in the system's local topology at its nonlinear eigenenergy relative to the system's topology in the absence of that state.
%
%
The topological robustness of the nonlinear mode is then related to the nonlinearly induced topological interface formed, and its existence is guaranteed as long as
\begin{equation}
\label{eq:nl_topo_robustness}
\norm{\Delta H(w)} \le \mu^\text{C}_\text{NL}
,
\end{equation}
where 
\begin{equation}
\mu^\text{C}_\text{NL} = \max_{\mathbf{x}} \big[ \mu^\text{C}_{(\mathbf{x},E_\text{NL})}(\mathbf{X},H_\text{NL}(\boldsymbol{\uppsi}_{\textrm{NL}})) \big]
\end{equation}
is the nonlinear localizer gap, namely the maximum localizer gap inside the newly created topological domain.
Physically, the robustness of the topological nonlinear mode can be interpreted as a guarantee to find a solution curve of the topological nonlinear mode over finite perturbation strength $w$ as long as the system perturbation is too weak to close the nonlinearly induced local gap. 

Probing the occupied nonlinear system's topology at an energy $E \ne E_\text{NL}$ can also provide useful information. 
In particular, the presence of a nonlinear mode can change the topology of the system over a range of energies, and thus may have use in a pump-probe-like setting. 
For example, a strong pump in a nonlinear photonic system may induce the appearance of topological interfaces at a wide range of frequencies, allowing for a weak signal at a different frequency from the pump to be routed to a particular outcoupling channel.


\subsection{Numerical $K$-theory}

One of the main benefits of the real-space spectral localizer approach is that it yields relatively simple formulae for a system's topological invariants and protection. 
This is because the underlying mathematics that determine the invariants of the dimensionally reduced system in 0D or 1D can be understood using only elementary homotopy theory (e.g.\ the fact that the classic group $\textup{GL}(n,\mathbb{R})$ has two connected components that are differentiated by the sign of the determinant). 
The full machinery of $K$-theory in the spectral localizer approach is hidden within the theorems that dictate how it performs dimensional reduction consistent with Bott periodicity, and is what guarantees that the topology of the dimensionally reduced effective system determines the topology of the original physical system. 
One consequence of this mathematical simplicity is that it results in numerically tractable formulae to determine a material's topology in any of the discrete symmetry classes and in any physical dimension --- by avoiding spectral flattening operations, the spectral localizer applied to sparse matrices remains sparse. 
This approach is in stark contrast to typical formulae derived using standard $K$-theory, which do not lend themselves to simple numerical implementations, nor efficient algorithms. 
As such, we refer to the present approach as numerical $K$-theory.

More broadly, numerical $K$-theory is the study of numerical algorithms to compute global or local $K$-theory invariants for matrix models of physical systems~\cite{Varjas2020, Fulga2012, Hastings2011, Leung2012}. 
Here we use the term numerical in the applied-math sense of algorithms that use floating-point arithmetic. 
Moreover, numerical $K$-theory methods have been successfully applied to large systems~\cite{Franca2023}, including real systems described by differential equations~\cite{Cerjan2022a}. 
Due to their relative simplicity, these methods can also inspire the development of novel experimental techniques allowing topological invariants derived through a numerical $K$-theory approach to be physically observed~\cite{Cheng2023}.



\subsection{Probing topological profile in nonlinear systems}

\begin{figure*}[t!]
\center
\includegraphics[width=2\columnwidth]{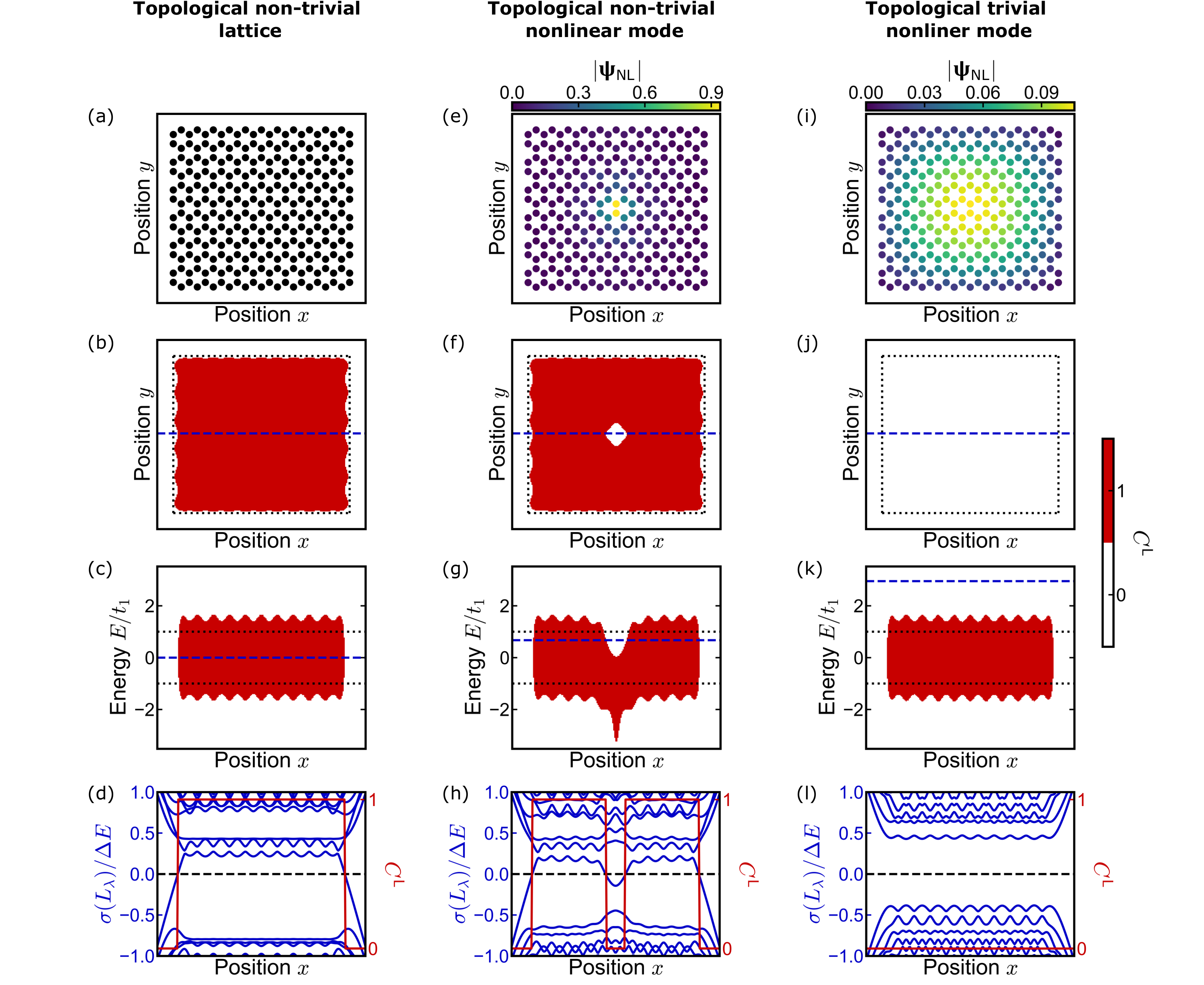}
\caption{
\textbf{Topological nonlinear modes.}
(a) Schematic of the geometry of the Haldane (honeycomb) lattice considered.
Local Chern number (b) $C_{(x,y,E_0)}^{\textrm{L}}$ and (c) $C_{(x,y_0,E)}^{\textrm{L}}$, where the white (red) regions stand for $C^{\textrm{L}}=0$ ($C^{\textrm{L}}=1$).
The blue dashed lines in (b) and (c) correspond to $y_0$ and $E_0$, respectively.
The black dotted line in (b) is a guide to the eye for the finite geometry of the lattice shown in (a).
The black dotted line in (c) illustrates the bulk band gap for a stripe geometry along the $x$-direction.
(d) Localizer spectrum $\sigma(L_{\boldsymbol{\lambda}=(x,y_0,E_0}))$ and local Chern number $C_{(x,y_0,E_0)}^{\textrm{L}}$ along the blue dashed line in (b).
(e) Field profile of the topological non-trivial nonlinear mode $|\boldsymbol{\uppsi}_\text{NL}|$.
(f)-(h) Same as (b)-(d), but for the topological non-trivial nonlinear mode in (e).
(i)-(l) Same as (e)-(h), except this is for a topological trivial nonlinear mode.
The parameters for the nonlinear Haldane model are chosen such that the lattice constant $a=1 [\text{a.u.}]$, the (next-)nearest-neighbor coupling $t_1 = 1 [\text{a.u.}]$ ($t_2/t_1 = 1/2$), inversion-symmetry-breaking mass term $m/t_1=0$, Haldane phase $\phi = \pi/2$, and $g = -2$, and $\kappa = 2 [t_1/a]$ has been set for all the localizer-related calculations.
$\Delta E$ stands for the bulk band gap at the honeycomb lattice's high-symmetry $\mathbf{K}$-point, $\Delta E = |m - 3 \sqrt{3} t_2 \sin(\phi)|$.
}
\label{fig:topo_nl_mode}
\end{figure*}

To illustrate the spectral localizer framework for classifying topology in nonlinear materials, we consider a system that can exhibit Chern insulating phases (i.e., a 2D class A system in the Altland-Zirnbauer classes~\cite{Altland1997}). 
In particular, we study a finite nonlinear Haldane lattice~\cite{Haldane1988} with the geometry shown in Fig.~\ref{fig:topo_nl_mode}(a) and with open boundary conditions. 
The lattice is characterized by a (next-)nearest neighbor coupling $t_1$ $(t_2/t_1 = 0.5)$, an inversion-symmetry-breaking on-site term set here to zero $m/t_1 = 0$, a time-reversal-symmetry-breaking Haldane flux $\phi=\pi/2$, and an on-site Kerr term $g |\uppsi_n|^2$ with $g$ being the nonlinear coefficient and $|\uppsi_n|^2$ being the intensity at the site $n$. 
The nonlinear Hamiltonian matrix $H_\text{NL}$ then reads
\begin{equation}
\label{eq:H_nl_e0}
[H_\text{NL}(\boldsymbol{\uppsi})]_{nl} = [H_0]_{nl} + g |\uppsi_n|^2 \delta_{nl}
, 
\end{equation}
with $H_0$ the linear Hamiltonian matrix for the Haldane lattice, and $\delta_{nl}$ the Kronecker delta function.

In the absence of any occupation, the linear lattice $(g=0)$ is in a topologically non-trivial phase because it satisfies the condition $|m/t_2| < 3\sqrt{3}|\sin(\phi)|$~\cite{Haldane1988}, and its bulk band gap has a Chern number $C=1$.
Figures~\ref{fig:topo_nl_mode}(b),(c) show the local Chern number [Eq.~\eqref{eq:local_chern_nb}] in position and energy $C_{(x,y,E)}^{\textrm{L}}$, directly revealing the local topological picture of this lattice, Fig.~\ref{fig:topo_nl_mode}(a).
As expected from topological band theory, one can see a non-trivial local Chern number $C^{\textrm{L}}=1$ inside the lattice region, delimited by the black dotted line, for choices of $E$ within the system's bulk band gap [Fig.~\ref{fig:topo_nl_mode}(b)]. 
Moreover, the local Chern number can also resolve the spectral extent of the system's non-trivial topology.
Indeed, Figure~\ref{fig:topo_nl_mode}(c) shows the local Chern number at a fixed $y$-coordinate position [see blue dashed line in Fig.~\ref{fig:topo_nl_mode}(b)], demonstrating in accordance to band theory that the system possesses a topological non-trivial energy range delimited by the bulk band gap found from a stripe geometry [see black dotted lines in Fig.~\ref{fig:topo_nl_mode}(b)].
Figure~\ref{fig:topo_nl_mode}(d) shows the localizer's spectrum $\sigma(L_{(x,y,E)})$ as $x$ is varied across the lattice for given $(y,E)=(y_0,E_0)$ [see blue dashed line in Fig.~\ref{fig:topo_nl_mode}(b) and Fig.~\ref{fig:topo_nl_mode}(c)]. 
This spectral flow demonstrates that, as the position is varied across the lattice's boundary, the local topological marker changes from trivial to non-trivial (or in reverse), which simultaneously forces the local gap to close, indicating the presence of a boundary-localized state.  
%
%


\subsubsection{Topological nonlinear modes}

With the inclusion of the model system's nonlinear response ($g \ne 0$), the model can be used to illustrate and distinguish nonlinear modes that are topological and trivial. 
To do so, we consider two different nonlinear modes and corresponding nonlinear eigenenergies that are found by self-consistently solving Eq.~\eqref{eq:nl_eq} [see Methods~\ref{sect:methods}]. 
In each case, the nonlinear Hamiltonian accounting for the system's occupation $H(\boldsymbol{\uppsi}_\text{NL})$ is then used to calculate the spectral localizer [Eq.~\eqref{eq:localizer}] and the associated local topological invariant and local gap at the nonlinear energy $E_\text{NL}$. 
For the case of the nonlinear mode and corresponding local gap and local invariant shown in Figs.~\ref{fig:topo_nl_mode}(e)-(h), the presence of the state in the system yields a change in the local topology where the state is localized and at its nonlinear eigenenergy, resulting in a nonlinearly induced topological interface. 
As such, this is a topological nonlinear mode. 
In contrast, the local gap and local invariant for the nonlinear mode in Figs.~\ref{fig:topo_nl_mode}(i)-(l) does not change the system's local topology at its non-linear eigenenergy, and as such is a trivial nonlinear mode.

Probing the occupied nonlinear system's topology at other energies provides additional insight, as shown in Fig.~\ref{fig:topo_nl_mode}(g).
For some $x$-positions near the nonlinear mode's center, the nonlinear mode shifts the energy range with non-trivial local topology to be lower, creating a trivial energy range inside the linear system's bulk band gap and a non-trivial energy range deep within the linear system's lower band.
Therefore, at the $(y,E)$-coordinates given by the blue dashed lines in Fig.~\ref{fig:topo_nl_mode}(f) and Fig.~\ref{fig:topo_nl_mode}(g), the localizer spectrum crosses the zero eigenvalue several times [Fig.~\ref{fig:topo_nl_mode}(h)].
To observe the shifted energy range due to the nonlinear mode's presence in the system, one would need to use a pump-probe-type experiment, pumping the system to create the intense, stationary nonlinear mode, and then using a weak signal at a different energy to probe the system and observe the nonlinearly induced topological interface.

We note that it may be possible for a trivial nonlinear mode, which does not induce a topological interface at its own nonlinear eigenenergy, to nevertheless shift the energy range of a system's topology similar to what is numerically observed in Fig.~\ref{fig:topo_nl_mode}(g). 
Thus, a trivial nonlinear mode may still have topologically nontrivial effects on a system.



\subsubsection{Robustness of topological nonlinear modes}

\begin{figure}[t!]
\center
\includegraphics[width=\columnwidth]{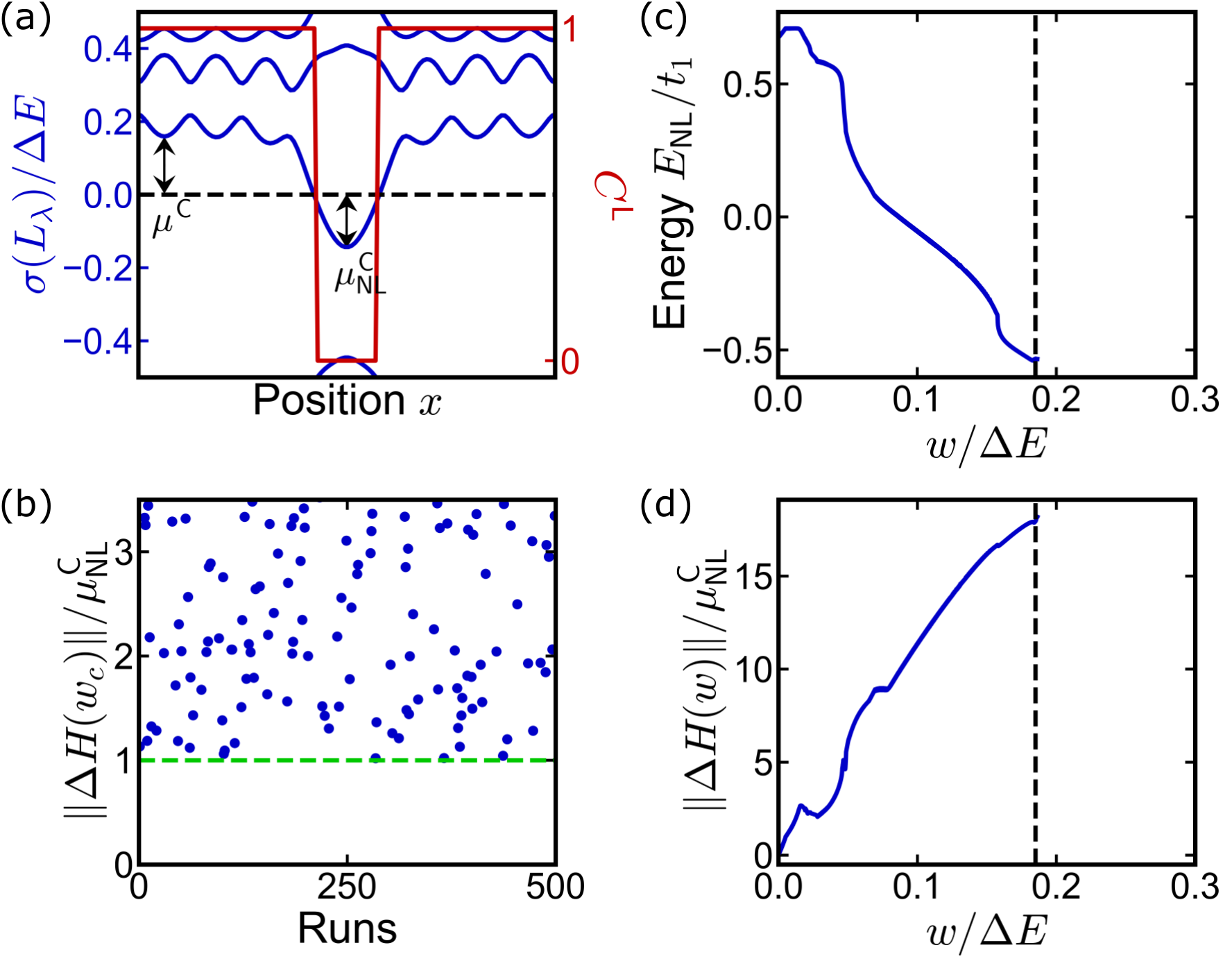}
\caption{
\textbf{Topological robustness of nonlinear modes.}
(a) Zoom-in of Fig.~\ref{fig:topo_nl_mode}(h) near the center of the nonlinear mode.
$\mu^\text{C}$ and $\mu^\text{C}_\text{NL}$ correspond to the linear and nonlinear localizer gaps.
(b) Norm of the maximum change of matrix Hamiltonian $\norm{\Delta H(w_c)}$, at the critical perturbation strength $w_c$, over several perturbation configuration runs.
(c),(d) Evolution of the nonlinear energy $E_\text{NL}$ and the change of matrix Hamiltonian $\norm{\Delta H(w)}$, respectively, against the perturbation strength $w$ for a given perturbation configuration.
The vertical black dashed lines depict the critical perturbation strength $w_c$ for the given perturbation configuration.
The parameters are the same as in Fig.~\ref{fig:topo_nl_mode}, and $w_f = 0.3 \Delta E$. 
}
\label{fig:topo_nl_robustness}
\end{figure}
%

%
The robustness of a topological nonlinear mode is characterized by its nonlinear localizer gap corresponding to its nonlinearly induced topological interface [Eq.~\eqref{eq:nl_topo_robustness}].
As discussed in Sect.~\ref{sect:loclaizer}, there are two types of local gaps that can be defined for nonlinear modes, $\mu^\text{C}$ and $\mu^\text{C}_\text{NL}$ [Fig.~\ref{fig:topo_nl_robustness}(a)], which stem from different properties of the system. 
The former ($\mu^\text{C}$) is an indication of the topological robustness of the underlying linear lattice, whereas the the latter ($\mu^\text{C}_\text{NL}$) indicates the nonlinear topological robustness associated with the nonlinear response due to the system's occupation.

The topological protection of a nonlinear mode can be interpreted as a guarantee to find its solution curve over a finite range of disorder strengths $w$ [Eq.~\eqref{eq:H_pert}] as long as Eq.~\eqref{eq:nl_topo_robustness} is satisfied.
To demonstrate how the local gap protects the existence of a nonlinear solution, we begin with the unperturbed topological nonlinear mode $\boldsymbol{\uppsi}_0$, obtained here in Fig.~\ref{fig:topo_nl_mode}(e), and calculate its solution curve using our nonlinear solver as we turn on the perturbation strength $w = 0 \rightarrow w_f$ [see Methods~\ref{sect:methods}].
Each point on the solution curve is calculated using the previous point as an initial guess [see Supplementary material~\cite{supp} for additional information].
During this procedure, either $w=w_f$ is reached; or after some finite value $w=w_c$, the nonlinear solver is not able to continue the solution curve, namely the solver is not converging given the previous nonlinear solution with $w<w_f$ used as an initial guess.
As we simulate an ensemble of hundreds of perturbation configurations that include on-site mass-like, $t_1$-like, and $t_2$-like perturbation terms drawn from a uniform distribution $[-w/2, w/2]$, we fail to converge to a solution only when $\norm{\Delta H(w_c)} > \mu^\textrm{C}_\textrm{NL}$ [Fig.~\ref{fig:topo_nl_robustness}(b)].
Figure~\ref{fig:topo_nl_robustness}(c)~and~\ref{fig:topo_nl_robustness}(d) give a example over a single perturbation configuration of the solution curve in the $(w, E_\text{NL})$ and $(w, \norm{\Delta H(w)})$ spaces, respectively.
Altogether, these simulations justify the claim that the topological robustness of a nonlinear mode guarantees the existence of a similar nonlinear eigenstate of the perturbed system.


\subsection{Probing topological dynamics in nonlinear systems}

The spectral localizer is capable of directly resolving topological dynamics of nonlinear systems. 
As the nonlinear spectral localizer is a pseudospectral approach that simultaneously accounts for both the system's spatial and energy information, its local markers can assess the system's topology while accurately incorporating any spatial inhomogeneities in the nonlinear system's occupation during the system's time-evolution.
In contrast, it is not possible to accurately assess such topological dynamics using band theory, as band theory requires assuming that the nonlinear system's occupation is effectively infinite so that the full system is periodic, resulting in a spatial averaging of the system's evolution.
Moreover, any complete theory of nonlinear topology must include a mechanism to resolve topological dynamics, as nonlinear systems are famously known to exhibit a wide range of dynamical behaviors.
For example, depending on the parameters of the system and the initial conditions, a nonlinear system's dynamics can cross a bifurcation point where its evolution can qualitatively change to become stable, periodic or even chaotic if slightly perturbed~\cite{Strogatz}.

To illuminate how the non-linear spectral localizer can be used to ascertain a system's nonlinear topological dynamics, we consider the general nonlinear rate equation with a Gaussian source $s_\text{in}(t)$ coupled to the system with coupling coefficient $\eta_\text{S}$ at the $n_\text{S}$-th site
\begin{equation}
\label{eq:rate_eq}
i \frac{d}{dt} \uppsi_n = \sum_l^N [H_\text{NL}(\boldsymbol{\uppsi})]_{nl} \uppsi_l + \eta_\text{S} s_\text{in} \delta_{l, n_\text{S}}
,
\end{equation}
where $N$ is the total number of sites in the lattice.
The rate equation [Eq.~\eqref{eq:rate_eq}] is integrated using a fourth-order Runge-Kutta method with $\uppsi_n = 0$, $\forall n$ as initial condition, and the Gaussian source given by
\begin{equation}
s_\text{in}(t) = s_0 e^{i \omega t} e^{-\tfrac{(t - t_0)^2}{2 \tau^2}}
,
\end{equation}
with $s_0$ the source amplitude, $\omega$ the source frequency, and $t_0$ and $\tau$ the temporal center and width, respectively, of the source.
In particular, the nonlinear spectral localizer's temporal analysis is demonstrated using two systems built from a Haldane model, the first of which includes a saturable nonlinear term on the inversion-breaking mass term $m$, while the second includes a Kerr-like term on the next-nearest neighbor coupling $t_2$. 
%


\subsubsection{Self-sustained topological nonlinear moving modes \label{sect:dynam1}}

\begin{figure*}[t!]
\center
\includegraphics[width=2\columnwidth]{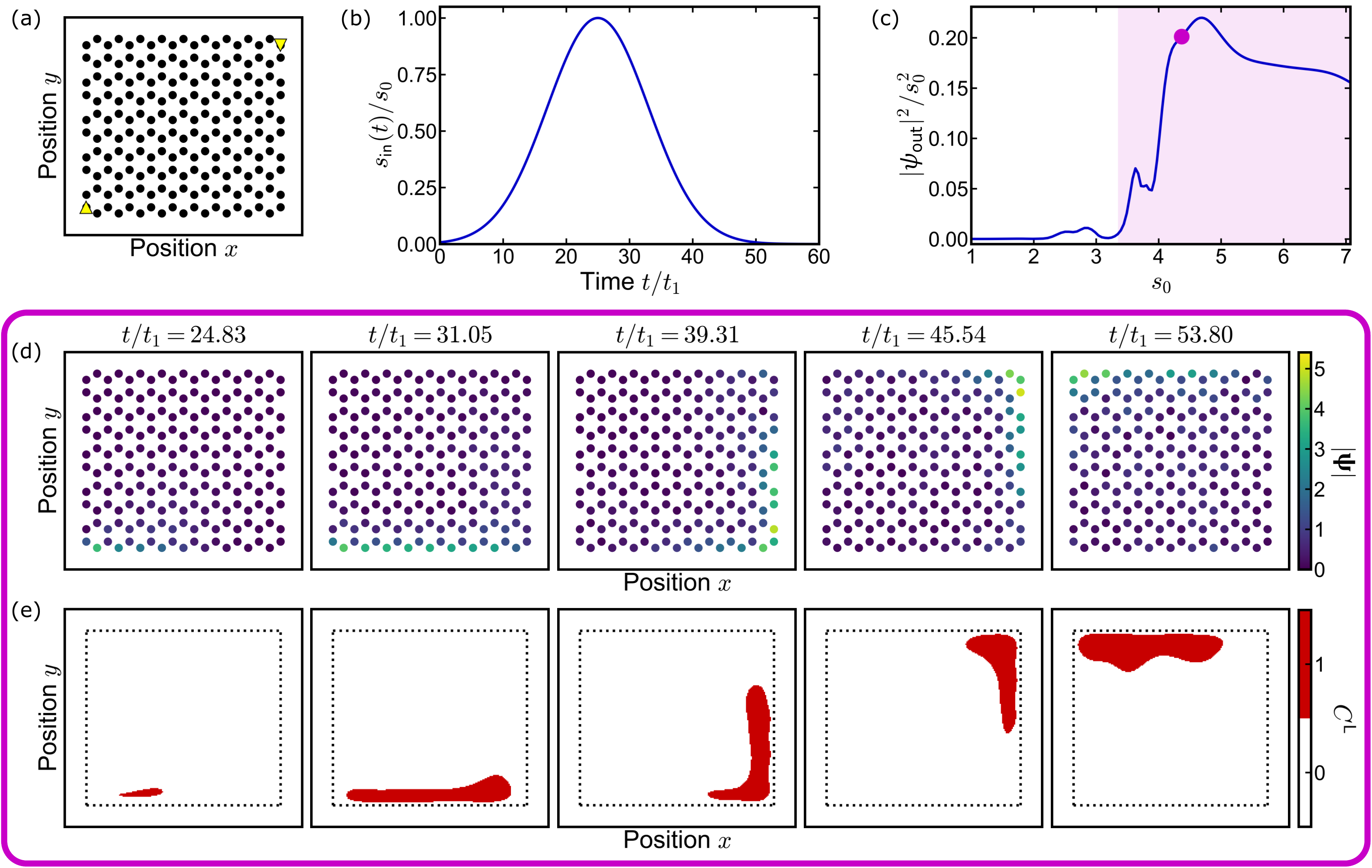}
\caption{
\textbf{Dynamics of a nonlinear moving mode in a saturable lattice.}
(a) Schematic of the lattice considered. 
The upward and downward yellow triangles depict the position of the Gaussian source (upward) and the position where the state's amplitude is captured (downward), namely the complex-valued amplitude $\uppsi_\text{out}$ at that site.
(b) Temporal evolution of the (normalized) Gaussian source $s_\text{in}$, with a source frequency $\omega/t_1 = 0$, a temporal center $t_0/t_1 = 25$, and temporal width $\tau/t_1 = 8$.
(c) Normalized intensity of the propagating mode at the output site $\uppsi_\text{out}$ against the source amplitude $s_0$, with the white (magenta) region illustrating the topological trivial (non-trivial) regime.
The output site intensity is captured at time $t/t_1=46.54$.
Snapshots over time of (d) the excited state $|\boldsymbol{\uppsi}|$ and (e) the local Chern number $C_{(x,y,0)}^{\textrm{L}}$ in real-space, given a source amplitude $s_0 = 4.4$ [magenta dot in (c)].
In (e), the black dotted lines are a guide-to-the-eye for the lattice geometry and the white (red) regions stand for $C^{\textrm{L}}=0$ ($C^{\textrm{L}}=1$).
The parameters for the saturable Haldane model are chosen such that the lattice constant $a=1 [\text{a.u.}]$, the (next-)nearest-neighbor coupling $t_1 = 1 [\text{a.u.}]$ ($t_2/t_1 = 1/3$), inversion-symmetry-breaking mass term $m/t_1=0$, Haldane phase $\phi = \pi/2$, and $m_0 = 2$, and $\kappa = 1 [t_1/a]$ has been set for all the local Chern number calculations.
The rate integration has been solved with a time-step $dt = 0.001$ and coupling coefficient $\eta_\text{S}/t_1 = 1$.
}
\label{fig:fig_dynamics_M_sat0_trivial}
\end{figure*}

As a first example, we look at a phenomenon where a self-sustained moving topological nonlinear mode is excited~\cite{Leykam2016, Zhou2017, Mukherjee2021}.
This effect is illustrated by considering a linear Haldane lattice in its trivial phase, with added nonlinearities that can locally drive the system into a topological phase. 
The system is realized by using state-dependent inversion-symmetry-breaking mass terms~\cite{Zhou2017}
\begin{equation}
m_n^{(A)}(\uppsi_n) = \frac{m_0}{1 + \gamma |\uppsi_n|^2}
, \quad
m_n^{(B)}(\uppsi_n) = \frac{-m_0}{1 + \gamma |\uppsi_n|^2}
,
\end{equation}
where $m_0$ is a reference inversion-symmetry-breaking mass term, $\uppsi_n$ is the complex valued amplitudes on the $n$-th lattice site (which could be either on the $A$ or $B$ sublattice), and $\gamma$ is the saturation coefficient.
The nonlinear Hamiltonian matrix $H_\text{NL}$ is then written as
\begin{equation}
[H_\text{NL}(\boldsymbol{\uppsi})]_{nl} = [H_0]_{nl} + m_n(\uppsi_n) \delta_{nl}
,
\end{equation}
with $m_n(\uppsi_n)$ being either $m_n^{(A)}(\uppsi_n)$ or $m_n^{(B)}(\uppsi_n)$ if the $n$-th site is in the $A$ or $B$ sublattice, respectively.

Inserting a narrow-frequency signal at the boundary of this saturable Haldane model, with a frequency within the linear lattice's bulk band gap, results in an edge state that remains localized to, and propagates only along, the system's boundary, as shown in Fig.~\ref{fig:fig_dynamics_M_sat0_trivial}. 
This phenomenon has been previously identified as being a self-sustained topological nonlinear moving mode~\cite{Leykam2016, Zhou2017, Mukherjee2021}, and it it was proposed that a topological phase transition occurs in the system if the source's amplitude $s_0$ is sufficiently large. 
However, previously, the topological phase transition could only be qualitatively explained using topological band theory, which required the assumption that the moving mode could be expanded to fill an entire, infinite lattice, to meet the necessary periodicity requirements for applying Bloch's theorem~\cite{Leykam2016, Zhou2017, Mukherjee2021}.

Instead, the local topology underlying the phenomena of self-sustained topological nonlinear moving modes can be directly captured in time using the nonlinear spectral localizer framework.
In particular, real-space snapshots of the topological dynamics quantitatively prove that the nonlinear mode's presence forces the system into a topological phase with a non-zero local Chern number in its vicinity [Figs.~\ref{fig:fig_dynamics_M_sat0_trivial}(d),(e)].
Moreover, the topologically non-trivial domain dynamically follows the nonlinear moving mode as it propagates around the lattice's boundary. 
Finally, the nonlinear spectral localizer can quantitatively confirm that a topological non-trivial region is only created when the source's intensity is high enough [Fig.~\ref{fig:fig_dynamics_M_sat0_trivial}(c)].
%


\subsubsection{Self-induced topological transition}

\begin{figure*}[t!]
\center
\includegraphics[width=2\columnwidth]{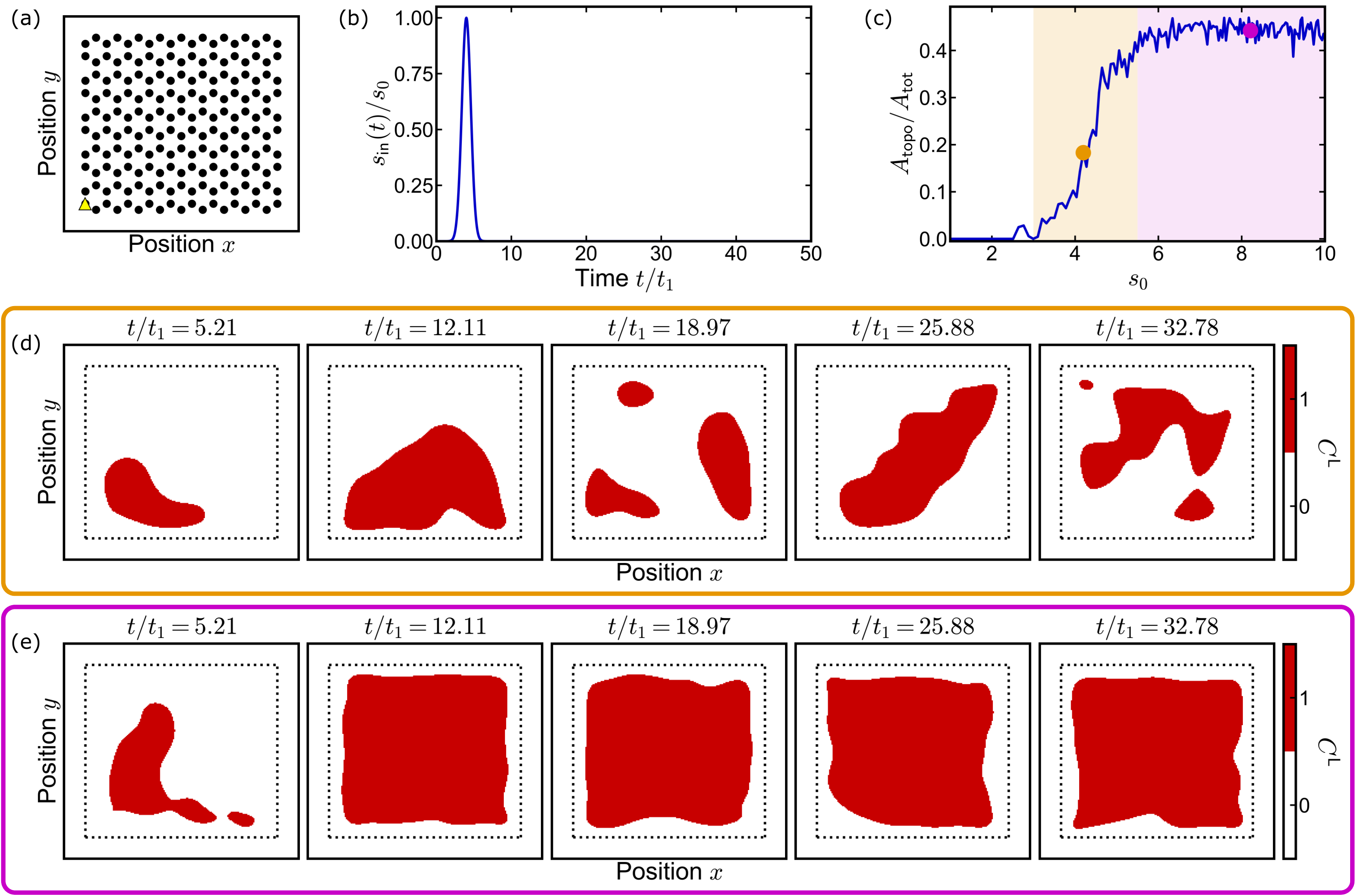}
\caption{
\textbf{Nonlinear dynamical regimes from a topological point-of-view.}
(a) Schematic of the lattice considered. 
The upward yellow triangles depict the position of the Gaussian source, with a source frequency $\omega/t_1 = 0$, a temporal center $t_0/t_1 = 4$, and temporal width $\tau/t_1 = 0.6$.
(b) Temporal evolution of the (normalized) Gaussian source $s_\text{in}$.
(c) Area of the lattice with of non-trivial local topology $A_{\textrm{topo}}$, normalized over the total area $A_\text{tot}$, of the system considered (including the trivial region outside the lattice), against the source amplitude $s_0$. The white, orange and magenta regions illustrate the topologically trivial, partly non-trivial and mostly non-trivial regimes, respectively.
The curve is plotted at time $t/t_1=50$.
Snapshot over time of the local Chern number $C_{(x,y,0)}^{\textrm{L}}$ for a source amplitude (d) $s_0 = 5.8$ [orange dot in (c)] and (e) $s_0 = 9.2$ [magenta dot in (c)].
The black dotted lines are a guide-to-the-eye for the lattice geometry and the white (red) regions denote $C^{\textrm{L}}=0$ ($C^{\textrm{L}}=1$).
The parameters for the nonlinear Haldane model are chosen such that the lattice constant $a=1 [\text{a.u.}]$, the (next-)nearest-neighbor coupling $t_1 = 1 [\text{a.u.}]$ ($t_2/t_1 = 1/3$), inversion-symmetry-breaking mass term $m/t_1 = 2$, Haldane phase $\phi = \pi/2$, and $\kappa = 1 [t_1/a]$ has been set for all the local Chern number calculations.
The rate integration has been solved with a time-step $dt = 0.0005$, coupling coefficient $\eta_\text{S}/t_1 = 1$.
}
\label{fig:fig_dynamics_t2_trivial}
\end{figure*}

As a second example for probing the dynamics of the nonlinear topology, we investigate the effect of the excitation source on self-inducing a topological phase transition across a full, finite system.
To do so, we consider a Haldane lattice that is topologically trivial in the linear regime with added Kerr-type nonlinearities on the next-nearest neighbor couplings, and excited by a spectrally broad Gaussian source, see Figs.~\ref{fig:fig_dynamics_t2_trivial}(a),(b).
The full nonlinear Hamiltonian for this system is
\begin{equation}
\label{eq:kerr_t2}
[H_\text{NL}(\boldsymbol{\uppsi})]_{nl} = [H_0]_{nl} + g \sum_{l \in \llangle n \rrangle} \left( |\uppsi_n|^2 + |\uppsi_l|^2 \right) e^{i \phi_{nl}}
,
\end{equation}
where $\llangle n \rrangle$ indicates the next-nearest neighbors to the $n$-th lattice site and $\phi_{nl}$ is Haldane phase for the couplings from site $l$ to $n$~\cite{Haldane1988}.
Notably, this example is distinguished from the previous example in Sect.~\ref{sect:dynam1} because the model in Eq.~\eqref{eq:kerr_t2} considers longer range nonlinearities; the local effect range is not on the single site where the state is, but on a longer range due to the nonlinear next-nearest-neighbor interaction.

Using the nonlinear spectral localizer framework, the long-range nonlinear Haldane model can exhibit a range of different regimes with distinct topological dynamics.
Depending on the excitation source amplitude $s_0$, the source partly or fully forces the unoccupied trivial lattice into a topologically non-trivial phase after a transient regime [Fig.~\ref{fig:fig_dynamics_t2_trivial}(c)].
These different dynamical regimes can be quantified using the area of the system with non-trivial local topology $A_{\textrm{topo}}$, normalized using the total lattice area $A_\text{tot}$, of the system considered (including the trivial region outside the lattice).
If the source amplitude is too small, the injected power to the lattice is not sufficient to induce a topological phase transition.
Increasing the source amplitude only partly changes the lattice's local topology, and realizes a dynamic topological regime where the location(s) of the non-trivial topology evolve in time [Fig.~\ref{fig:fig_dynamics_t2_trivial}(d)].
Finally, when the source is strong enough, most the of the lattice is forced into a topologically non-trivial phase [Fig.~\ref{fig:fig_dynamics_t2_trivial}(e)].
This model system exemplifies how a nonlinear system can be used to dynamically create topological interfaces, as well as to understand the lifetimes of these interfaces.


\section{Conclusions}

In conclusion, we have developed a general framework, based on numerical $K$-theory, for classifying topology in nonlinear topological insulators.
Using variants of the Gross-Pitaevskii equation~\cite{Gross1961, Pitaevskii1961}, we have demonstrated the ability of the proposed framework to capture the topological landscape in both real-space and energy, as well as the topological dynamics of the system.
In doing so, we have rigorously studied and mathematically proven a number of qualitative claims present in the literature of nonlinear topological insulators~\cite{Smirnova2020}.
Indeed, given the nonlinear spectral localizer's ability to provide a quantitative definition of nonlinear topological modes based on their ability to induce a topological interface, we can directly determine these modes' topological robustness, and have shown that this topological protection guarantees a nearby nonlinear eigenmode solution to the nonlinear Hamiltonian.
Moreover, we have demonstrated how this approach enables the study of topological dynamics within nonlinear topological insulators, where different dynamical regimes can be obtained depending on the source amplitude.
Looking forward, we anticipate that the nonlinear spectral localizer can be used to design systems whose topology can be dynamically controlled in more sophisticated arrangements that involve energy transitions and higher-order nonlinear processes such as multi-wave mixing~\cite{Pilozzi2017, Zhang2019, Mittal2021, Jia2023}, and that potentially yield topological pump-probe experiments. Moreover, as the linear spectral localizer can be applied to aperiodic systems~\cite{Fulga2016, Jia2023}, non-Hermitian systems~\cite{Cerjan2023}, and realistic photonic crystals~\cite{Cerjan2022a}, the nonlinear spectral localizer should be able to predict nonlinear topological modes and topological dynamics across a broad range of materials and experimentally realizable platforms.


\section*{Methods}
\label{sect:methods}

For the sake of completeness, we provide the details of the methods utilized to solve the nonlinear eigenvalue equation [Eq.~\eqref{eq:nl_eq} in the main text] with 
$H_\text{NL}$ the $(N \times N)$ nonlinear Hamiltonian matrix, which is explicitly composed of a linear part $H_0$ and a nonlinear part $H_1 (|\boldsymbol{\uppsi}_\text{NL}|^2)$,
$\boldsymbol{\uppsi}_\text{NL}$ the nonlinear $(N \times 1)$ eigenvector, and
$E_\text{NL}$ the nonlinear eigenvalue, 
where $N$ is the total number of sites.
Particularly , we mainly used variant of the function \textbf{fsolve} in Matlab~\cite{MATLAB} that uses a specific gauge and a power constraint, and a discrete version of the Petviashvili method~\cite{Petviashvili2016, Christiansen1996}.


\subsection{fsolve from Matlab}
\label{sect:fsolve}

The function \textbf{fsolve} in Matlab~\cite{MATLAB} is used to solve a system of nonlinear equations.
The problem is typically specified by a loss function $\boldsymbol{G}$ that \textbf{fsolve} tries to minimize given an initial guess of the unknowns.
For that purpose, Equation~\eqref{eq:nl_eq} is re-written as
\begin{equation}
\label{eq:nl_eq_bis}
\boldsymbol{F}(\boldsymbol{\uppsi}) = H_\text{NL}(\boldsymbol{\uppsi}) \boldsymbol{\uppsi} - E \boldsymbol{\uppsi} = \boldsymbol{0}
,
\end{equation}  
where $\boldsymbol{F}(\boldsymbol{\uppsi})$ is a $(N \times 1)$-vector.
For better convergence, Equation~\eqref{eq:nl_eq_bis} is separated into its real and imaginary part
\begin{equation}
\boldsymbol{G}(\boldsymbol{\uppsi}) = 
\left(
\begin{array}{c}
\text{Re} \left[ \boldsymbol{F}(\boldsymbol{\uppsi}) \right] \\[0.7ex]
\text{Im} \left[ \boldsymbol{F}(\boldsymbol{\uppsi}) \right]
\end{array} 
\right)
,
\end{equation}
and the problem is now specified by
\begin{equation}
\label{eq:fsolve_nl_eq}
\boldsymbol{G}(\boldsymbol{u}) = \boldsymbol{0}
\end{equation}
where
$\boldsymbol{G}$ is a $(2N \times 1)$-vector-valued function, and
$\boldsymbol{u} = (\ldots, \alpha_n, \ldots, \beta_n, \ldots, E) \equiv (\boldsymbol{\uppsi}, E)$ is a $(2N+1 \times 1)$-vector composed of the $2N+1$ unknowns, with $\uppsi_n = \alpha_n + i \beta_n$.

In order to greatly speed-up the computation, we provide to \textbf{fsolve} the Jacobian, $J$, of $\boldsymbol{G}$
\begin{equation}
J_{nl} = \frac{\partial G_n}{\partial u_l} 
,
\end{equation}
which, in term of terms of the derivatives of $\boldsymbol{F}$, explicitly reads
\begin{equation}
\frac{\partial \boldsymbol{G}}{\partial u_l} = 
\left(
\begin{array}{c}
\text{Re} \left[ \frac{\partial \boldsymbol{F}}{\partial u_l} \right] \\[1.5ex] 
\text{Im} \left[ \frac{\partial \boldsymbol{F}}{\partial u_l} \right]
\end{array} 
\right)
,
\end{equation}
where the derivatives of $\boldsymbol{F}$ are given by:
\begin{align}
\label{eq:jacobian_F_alpha}
\begin{split}
\frac{\partial F_n}{\partial \alpha_l} 
& = [H_{0}]_{nl} + g \left( 2 \alpha_n (\alpha_n + i \beta_n) + (\alpha_n^2 + \beta_n^2) \right) \delta_{nl} \\
& \quad - E \delta_{nl}
\end{split}
, \\
\label{eq:jacobian_F_beta}
\begin{split}
\frac{\partial F_n}{\partial \beta_l} 
& = i H_{nl} + g \left( 2 \beta_n (\alpha_n + i \beta_n) + i (\alpha_n^2 + \beta_n^2) \right) \delta_{nl} \\
& \quad - i E \delta_{nl}
\end{split}
, \\
\label{eq:jacobian_F_E}
\frac{\partial F_n}{\partial E} 
& = - (\alpha_n + i \beta_n)
.
\end{align}


\subsubsection{fsolve from Matlab, with gauge}
\label{sect:fsolve_gauge}

The problem specified by Eq.~\eqref{eq:fsolve_nl_eq} is actually not well defined because there is $2N+1$ unknowns for $2N$ equations.
To work out this issue, one can either provide an additional equation such as by constraining the power $\norm{\boldsymbol{\uppsi}}$ to some finite value, or simply to get rid off of one unknown.
We here choose the latter, namely we get rid off of one unknown by fixing the $U(1)$ gauge freedom of the system.
Indeed, if $\boldsymbol{u} = (\boldsymbol{\uppsi}, E)$ is a solution of the nonlinear equation Eq.~\eqref{eq:nl_eq}, then $\boldsymbol{\tilde{u}} = (\boldsymbol{\uppsi} e^{i\phi}, E)$ is also a solution because of the $|\boldsymbol{\uppsi}|^2$ term in $H_1$.
Besides, if we have Eq.~\eqref{eq:nl_eq_bis}, then we also have
\begin{equation}
H_\text{NL}(\boldsymbol{\uppsi}) \frac{\boldsymbol{\uppsi}}{Z} - E \frac{\boldsymbol{\uppsi}}{Z} = \boldsymbol{0}
,
\end{equation}
with $Z$ some constant.
We therefore decide to fix the gauge by defining $Z$ as
\begin{equation}
Z = \alpha_\nu
,
\end{equation}
with $\nu = \arg \max_{j=1,\ldots,N} \big[ |\text{Re}(\uppsi_j)| \big]  = \arg \max_{j=1,\ldots,N} \big[ |\alpha_j| \big]$.
This is equivalent as setting one real-part component to $1$, and thereby removing it from the set of unknowns.

Consequently, we solve a slightly modified problem specified by the ``gauged" function $\boldsymbol{\tilde{G}}$
\begin{equation}
\label{eq:fsolve_nl_eq_gauge}
\boldsymbol{\tilde{G}}(\boldsymbol{\tilde{u}}) = \boldsymbol{0}
\end{equation}
where
$\boldsymbol{\tilde{G}}$ is a $(2N \times 1)$-vector-valued function, and
$\boldsymbol{\tilde{u}} = (\ldots, \tilde{\alpha}_n, \ldots, \tilde{\alpha}_{\nu-1}, 1, \tilde{\alpha}_{\nu+1}, \ldots, \tilde{\beta}_n, \ldots, E)$ is a $(2N+1 \times 1)$-vector composed of the $2N$ unknowns, with $\tilde{\alpha}_n = \alpha_n/Z$ and $\tilde{\beta}_n = \beta_n/Z$.
Finally, the solution of the original equation [Eq.~\eqref{eq:nl_eq}] is retrieved by multiplying by $Z$
\begin{equation}
\boldsymbol{\uppsi} = Z \tilde{\boldsymbol{\uppsi}}
. 
\end{equation}

As we are solving for the ``gauged" function $\boldsymbol{\tilde{G}}$, the corresponding Jacobian is also slightly modified.
Nonetheless, the ``gauged" Jacobian can still be written in terms of of the ``ungauged" Jacobian, namely using the ungauged variables.
Indeed, we have
\begin{equation}
\boldsymbol{\tilde{F}}(\ldots, \tilde{\alpha}_n, \ldots, \tilde{\beta}_n, \ldots, E) = \frac{1}{Z} \boldsymbol{F}(\ldots, \alpha_n, \ldots, \beta_n, \ldots, E) 
,
\end{equation}
thereby:
\begin{align}
\label{eq:jacobian_F_alpha_gauge}
\frac{\partial \tilde{F}_n}{\partial \tilde{\alpha}_l}
& = \frac{\partial F_n}{\partial \alpha_l}
, \\[1.ex]
\label{eq:jacobian_F_beta_gauge}
\frac{\partial \tilde{F}_n}{\partial \tilde{\beta}_l}
& = \frac{\partial F_n}{\partial \beta_l}
, \\[1.ex]
\label{eq:jacobian_F_E_gauge}
\frac{\partial \tilde{F}_n}{\partial E}
& = \frac{1}{Z} \frac{\partial F_n}{\partial E}
.
\end{align}
%


\subsubsection{fsolve from Matlab, with gauge and constraint on power}
\label{sect:fsolve_gauge_P}

For some of the nonlinear modes obtained we also decided to add a constraint to the nonlinear solution, namely to fix the total power of the mode $P = \norm{\boldsymbol{\uppsi}}^2$ to some finite value.
Keeping the ``gauge" used previously, the additional power constraint is realized by changing Eq.~\eqref{eq:nl_eq_bis} as
\begin{equation}
H_\text{NL} \left( \sqrt{P} \frac{\boldsymbol{\uppsi}}{\norm{\boldsymbol{\uppsi}}} \right) \frac{\boldsymbol{\uppsi}}{Z} - E \frac{\boldsymbol{\uppsi}}{Z} = \boldsymbol{0}
,
\end{equation}
which gives
\begin{equation}
\label{eq:fsolve_nl_eq_gauge_constraint}
\boldsymbol{\tilde{G}}(P; \boldsymbol{\tilde{u}}) = \boldsymbol{0}
,
\end{equation}
where
$\boldsymbol{\tilde{G}}$ is a $(2N \times 1)$-vector-valued function, 
$P$ is the given parameter, and
$\boldsymbol{\tilde{u}} = (\ldots, \tilde{\alpha}_n, \ldots, \tilde{\alpha}_{\nu-1}, 1, \tilde{\alpha}_{\nu+1}, \ldots, \tilde{\beta}_n, \ldots, E)$ is a $(2N+1 \times 1)$-vector composed of the $2N$ unknowns, with $\tilde{\alpha}_n = \alpha_n/Z$ and $\tilde{\beta}_n = \beta_n/Z$.
The rest of the method proceeds as in Sect.~\ref{sect:fsolve_gauge}, except here we did not pass a Jacobian to \textbf{fsolve}.


\subsection{Petviashvili}
\label{sect:petviashvili}

The Petviashvili method is another method computing nonlinear solutions~\cite{Petviashvili2016, Christiansen1996}.
For the purpose of the Petviashvili method, Equation~\eqref{eq:nl_eq} is re-written as
\begin{equation}
\label{eq:petviashvili}
M \boldsymbol{\uppsi} = \boldsymbol{\uppsi}^{(p)}
,
\end{equation}
with
\begin{equation}
M = - (H_0 - E I)
,
\end{equation}
and
\begin{equation}
\label{eq:petviashvili_psi_p}
\boldsymbol{\uppsi}^{(p)} = H_1(|\boldsymbol{\uppsi}|^2) \boldsymbol{\uppsi}
.
\end{equation}
The $p$ superscript in $\boldsymbol{\uppsi}^{(p)}$ roughly stands for its polynomial degree, namely $p=3$ for the case of a Kerr-like term $\uppsi^{(p)}_n = |\uppsi_n|^2 \uppsi_n$, and $I$ is the identity matrix.

Given an initial guess $\boldsymbol{\uppsi}_0$, the Petviashvili method is an iterative method that computes
\begin{equation}
\label{eq:petviashvili_psi_np1}
\boldsymbol{\uppsi}_{k+1} = S_k^\gamma M^{-1} \boldsymbol{\uppsi}^{(p)}_k
,
\end{equation}
where 
$\gamma$ is a constant, 
$\boldsymbol{\uppsi}^{(p)}_k$ is $\boldsymbol{\uppsi}^{(p)}$ [Eq.~\eqref{eq:petviashvili_psi_p}] calculated with $\boldsymbol{\uppsi} = \boldsymbol{\uppsi}_k$, and
\begin{equation}
\label{eq:petviashvili_S_k}
S_k = \frac{\langle M \boldsymbol{\uppsi}_k, \boldsymbol{\uppsi}_k \rangle}{\langle \boldsymbol{\uppsi}^{(p)}_k, \boldsymbol{\uppsi}_k \rangle}
\end{equation}
with $\langle \boldsymbol{v}, \boldsymbol{w} \rangle = \sum v_i^* w_i$.
Then, the iteration stops whenever
\begin{equation}
\label{eq:petviashvili_tol}
\norm{\boldsymbol{\uppsi}^{(p)} - M \boldsymbol{\uppsi}_k)}_1 < \epsilon
,
\end{equation}
otherwise the iteration continues by replacing $\boldsymbol{\uppsi}_{k+1} \rightarrow \boldsymbol{\uppsi}_k$.
In Equation~\eqref{eq:petviashvili_psi_np1}, $\gamma$ is a constant that can be heuristically chosen as
\begin{equation}
\gamma = \frac{p}{p-1}
.
\end{equation}




\bibliography{ref_v0}

\section*{Acknowledgments}
A.C., T.A.L., and S.W.\ acknowledge support from the Laboratory Directed Research and Development program at Sandia National Laboratories. This work was performed in part at the Center for Integrated Nanotechnologies, an Office of Science User Facility operated for the U.S. Department of Energy (DOE) Office of Science.
Sandia National Laboratories is a multimission laboratory managed and operated by National Technology \& Engineering Solutions of Sandia, LLC, a wholly owned subsidiary of Honeywell International, Inc., for the U.S. DOE's National Nuclear Security Administration under Contract No. DE-NA-0003525. 
The views expressed in the article do not necessarily represent the views of the U.S. DOE or the United States Government.
T.A.L. acknowledges support from the National Science
Foundation, Grant No. DMS-2110398.

\end{document}